\pdfoutput=1

\documentclass[aps,prb,reprint,showpacs,superscriptaddress]{revtex4-1}

\usepackage{graphicx}
\usepackage{graphics}
\usepackage{amsmath}
\usepackage{amssymb}
\usepackage{amsfonts}
\usepackage{dcolumn}
\usepackage{dsfont}
\usepackage{latexsym}
\usepackage{rotating}
\usepackage{color}
\usepackage{latexsym}
\usepackage{bbm}
\usepackage{subfigure}
\usepackage{float}
\usepackage{epsfig}
\usepackage{psfrag}
\usepackage{bm}
\usepackage{amsthm}
\usepackage{eucal}
\usepackage{mathrsfs}
\usepackage{url}

\usepackage{color} 

\usepackage{hyperref}
\hypersetup{
colorlinks=true,final=true,
        linkcolor=blue,
        citecolor=blue,
        filecolor=blue,
        urlcolor=blue,
}
\begin{document}
\title{Broadband pump-probe study of biexcitons in chemically exfoliated layered WS$_{2}$}
\author{R. K. Chowdhury}
\author{S. Nandy}
\author{S. Bhattacharya}
\author{M. Karmakar}
\affiliation{Department of Physics, Indian Institute of Technology Kharagpur, Kharagpur - 721302, India}
\author{B. N. S. Bhaktha}
\author{P. K. Datta}
\affiliation{Department of Physics, Indian Institute of Technology Kharagpur, Kharagpur - 721302, India}
\author{A. Taraphder}
\affiliation{Department of Physics, Indian Institute of Technology Kharagpur, Kharagpur - 721302, India}
\affiliation{Centre for Theoretical Studies, Indian Institute of Technology Kharagpur, W.B. 721302, India}
\author{S. K. Ray}
\affiliation{Department of Physics, Indian Institute of Technology Kharagpur, Kharagpur - 721302, India}
\affiliation{S. N. Bose National Centre for Basic Sciences, Kolkata – 700106, India}

\begin{abstract}
Strong light-matter interactions in layered transition metal dichalcogenides (TMDs) open up vivid possibilities for novel exciton-based devices. The optical properties of TMDs are dominated mostly by the tightly bound excitons and more complex quasiparticles, the biexcitons.  Instead of physically exfoliated monolayers, the solvent-mediated chemical exfoliation of these 2D crystals is a cost-effective, large-scale production method suitable for real device applications. We explore the ultrafast excitonic processes in WS$_{2}$ dispersion using broadband femtosecond pump-probe spectroscopy at room temperature. We detect the biexcitons experimentally and calculate their binding energies, in excellent agreement with earlier theoretical predictions. Using many-body physics, we show that the excitons act like Weiner-Mott excitons and explain the origin of excitons via first-principles calculations. Our detailed time-resolved investigation provides ultrafast radiative and non-radiative lifetimes of excitons and biexcitons in WS$_{2}$. Indeed, our results demonstrate the potential for excitonic quasiparticle-controlled TMDs-based devices operating at room temperature.  
\end{abstract}
\maketitle
\section{Introduction}
Transition metal dichalcogenides (TMDs), playing a leading role in the development of two dimensional (2D) semiconducting physics, exhibit superior optical properties that are substantially different from their bulk contribution~\cite{Zenga tmd, Xu tmd}. In the family of 2D TMDs, one of the most extensively studied material is tungsten disulphide (WS$_{2}$) that displays a wide variety of intriguing properties such as layer-tunable band gap modulation, large spin-valley splitting, high absorption coefficient and ultrafast excitonic properties~\cite{Wang prop., Balandin prop., Novoselov prop., Kobayashi prop., Kim prop.}. Most importantly, one can tune the properties by varying the number of layers, which makes this material a promising candidate for novel device applications. Using variable range of architectures, WS$_{2}$ is being constantly explored for low-cost, sensitive and scalable optoelectronic devices like ultra-high responsive photosensors, valleytronics, and excitonic devices~\cite{Kwon application, Science application, app1, app2, app3}.

In general, atomically thin WS$_2$, consisting of two hexagonally bonded sulfur (S) atoms, bonded with transition metal tungsten (W), get stacked by weak van-der-Waals force of attraction analogous to graphene~\cite{Geim graphene exfoliation}. As a result of this weak interaction, 2D layered WS$_2$ can be exfoliated readily that results in layer-dependent tunable energy band structure resulting in direct band gap transition for the monolayer. The layered WS$_2$ also exhibits unique phenomenon called band nesting (BN), where the valance band (VB) and conduction band (CB) are parallel to each other along a specific k-path (towards $\Gamma$-K direction). Due to the BN effect, the photo-generated charge carriers with opposite polarities  move with equal but opposite velocities in the CB and VB, exhibiting extraordinary optical characteristics as  reported earlier~\cite{Carvalho BN}. Furthermore, unlike conventional semiconductors, Coulomb interaction in TMDs is remarkably robust because of the excited carrier screening and Pauli blocking~\cite{Schmidt ws2 pumpprobe, Sie MoS2 pump probe, Dielectric theory1, Dielectric theory2}. As a result, various excitonic quasiparticles like exciton, biexciton (two neutral excitons bound by interaction) are predicted to co-exist inside the exfoliated 2D stratum at room temperature~\cite{Aleithan MoS2 band map pumpprobe, Sim MoS2 exciton, Mak MoS2 trion, Zhu exciton WS2, Wang exciton, Paradisanos biexciton WS2, Yuan WS2 exciton, Biexciton B.E. expt wse2, Dark trion ws2}. The steady-state optical studies are incapable to reveal the nature of these excitonic phenomena and therefore, time-resolved optical studies are required to investigate the ultrafast carrier dynamics of layered WS$_2$.

On the other hand, as compared to physical exfoliation, chemical vapor deposition (CVD) and solvent-mediated chemical exfoliation techniques of 2D TMDs are contemporaneous in terms of wafer-scale production, though the last one appears to be cost-effective for device applications~\cite{Geim graphene exfoliation, Chowdhury exfoliation, Chowdhury exfoliation2}. Therefore, chemically exfoliated mono-to-few layer WS$_2$ has been widely used for the nanofabrication of various optoelectronic components~\cite{Borzda MoS2 chem pump probe, Tsokkou MoS2 solution pump probe}. However, the manifestation of the devices for futuristic applications requires a profound knowledge of the excitonic energy levels, formation and evolution of multiexcitonic states and their decay dynamics, which are still lacking~\cite{Vega-Mayoral charge dynamics WS2, Chernikov exciton BE WS2}. In the present work, we have chosen the time-resolved pump-probe spectroscopy of chemically exfoliated WS$_2$ layers to study the above ultrafast processes in detail.

We have explained the transient absorption spectroscopy (TAS) signals of chemically exfoliated mono-to-quad layer WS$_2$ flakes dispersed in dimethylformamide (DMF) solvent in the following. A horizontally polarized optical pump (405 nm, 3 eV) with varying power along with a broadband (350-750 nm) optical probe was used for the ultrafast pump-probe measurements. The Raman fingerprint have confirmed the exfoliation of mono-to-quad layer of WS$_2$ flakes in DMF. Using density functional theory (DFT) calculations, we have also shown that the degeneracy and splitting of VB at different k-points are always present, which are similar in nature specially at K and K$^{\prime}$ points in the presence of atomic spin-orbit coupling from mono-to-quad layer transformation. Our theoretical calculations suggest that the physical reasons for the generation and evolution of excitons, biexcitons and their corresponding decay dynamics do not change with increase in number of layers up to quadlayers. Furthermore, a detailed investigation of TAS signal of the layered WS$_2$ dispersion confirms six major features - (i) three saturation absorptions (SA) which are governed by steady state UV-Vis absorption and (ii) three excited state absorptions that substantiate the presence of excitonic quasiparticles like excitons and biexcitons. TAS spectra confirm that the relative oscillator strength of the deconvoluted biexcitonic peaks increases with probe delay with respect to the excited state absorption (ESA), which is discussed in terms of hot exciton cooling process. The binding energies of AA ($\sim$ 69 meV) and BB ($\sim$ 66 meV) biexcitons have been calculated experimentally from blue shifted biexcitonic peaks. We have also measured the power dependent red-shift of the biexcitonic peak, which is the result of many-body interactions govenred by excited state screening and reduction of excitonic band-gap due to band renormalization at higher pump excitations~\cite{Many body mos2, Ferrari Band renorm., Dielectric theory1, Dielectric theory2}. The ultrafast non-radiative and radiative lifetimes of the excitons and biexcitons have also been extracted using 3$^{rd}$ exponentially fitted transient absorption spectra.



\section{Methods}
\subsection{Sample synthesis} 
For the synthesis of WS$_{2}$, we have used the exfoliation technique mentioned in the earlier reports from our group~\cite{Chowdhury exfoliation, Arupda WS2}. Resulting flakes have a lateral dimension ranging from a few hundred nanometers to a micron with $\sim$1-4 layers WS$_{2}$ flakes dispersed uniformly in DMF solvent as confirmed by Raman analysis.

\subsection{Raman and steady state absorption characterization} 
Raman spectra of WS$_{2}$ on silicon were recorded using a T64000 (JobinYvon Horiba) spectrometer with an Argon-Krypton mixed ion gas laser (514.5 nm). The absorption spectra of the dispersion of layered WS$_{2}$ were acquired using a PerkinElmer Lambda-2 spectrometer in a wavelength range of 190–1100 nm.

\subsection{Transient absorption spectroscopy} Ultrafast transient absorption spectra of WS$_2$ dispersion is recorded with a commercially available transient absorption spectrometer (Newport) using 50 fs Ti: Sapphire mode locked amplifier (Libra-He, Coherent) with 808 nm center wavelength and repetition rate of 1 kHz. Major part (70$\%$) of the fundamental beam goes into optical parametric amplifier (TOPAS prime, Coherent), which acts as tunable pump source. The other part of the fundamental beam is focused in a 3 mm CaF$_2$ crystal which generates linearly polarized white light continuum (350-750 nm) probe. The latter one passes through an optical delay channel using a motorized translational stage in order to probe the system at different times after the pump excitation has perturbed the system. Pump and probe beam spot-sizes are maintained at 2 mm and 0.16 mm respectively which is used for this non-collinear pump-probe measurement setup. This kind of arrangements provide the pump fluence as 1.8 GW/cm$^{2}$ for 3 mW time averaged pump power. A spectrometer equipped with linear Si photodiode array is used to detect the differential probe spetra. Differential absorbance (${\Delta}$A) spectra are measured by modulating the pump beam at 500 Hz with an optical chopper which is again fed to the spectrometer.

\begin{figure}[htb]
\begin{center}
\epsfig{file=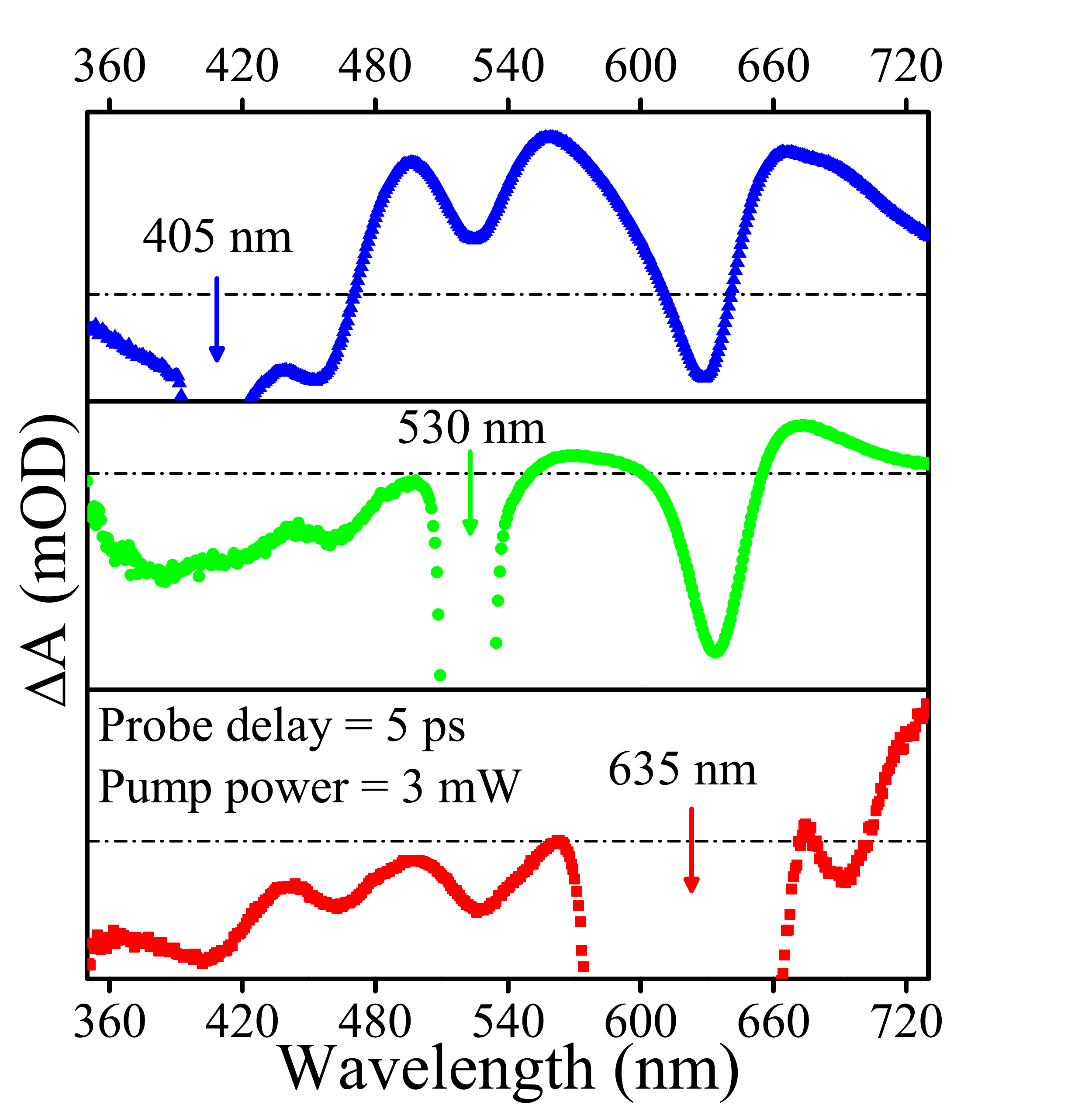,trim=0.0in 0.05in 0.0in 0.05in,clip=true, width=90mm}\vspace{0em}
\caption{(Color online) TAS spectra of mono-to-quad layer dispersion of WS$_{2}$, recorded at fixed probe delay t = 5 ps, for three pump excitation (3 mW) wavelengths, i.e., in resonance with the A ($\lambda_{pump}$ = 635 nm) and B ($\lambda_{pump}$ = 530 nm) excitons, and out of resonance with C ($\lambda_{pump}$ = 405 nm).}
\label{fig41}
\end{center}
\end{figure}

\subsection{Resonant and off-resonant pump excitation.} TAS spectra have been obtained for different pump excitation wavelengths starting from 405 nm, i.e. higher than the A, B and C excitonic wavelength, along with two other excitation (635 nm and 530 nm) resonant with A and B respectively (Fig.~\ref{fig41}). Interestingly, all the six major spectral features (SA and ESA) are still observed for the lower energy resonant pumping conditions. The SA at B for the excitation at A ($\lambda_{pump}$ = 635 nm) can be explained by extremely fast intervalley spin relaxation at K and K$^{\prime}$ points. The unexpected presence of the higher energy C proves that all the three excitons (A, B and C) are resonantly coupled with each other.  
   

\subsection{Computational Details} The electronic band structure calculations of monolayer, bilayer, trilayer and quadlayer WS$_{2}$ have been performed in the framework of ab initio density functional theory (DFT) in conjunction with all-electron projector augmented wave potentials and the Perdew-Burke-Ernzerhof generalized gradient approximation (PBE-GGA)~\cite{pbe} to the electronic exchange and correlation, as implemented in the WIEN2K code~\cite{wien}, which has been used successfully in various instances for describing low-dimensional systems. We incorporate the atomic spin orbit coupling with GGA calculation in terms of the second-variational method with scalar-relativistic orbitals as a basis. For all four cases, we have created a periodic slab geometry with a 20-$A^{\circ}$-thick vacuum layer along Z axis. We have relaxed both the in-plane lattice constant $\mathbf{a}$ and the positional parameter $\mathbf{z}$ until the force on every atom becomes less than 2 mRy/Bohr. We have taken the same values of $R_{mt} = 2.1$ a.u., $R_{mt}K_{max} =8$, and $l_{max} = 12$ and same exchange-correlation functional in each calculation. We have employed a k-mesh of $39 \times 39 \times 2$ for the monolayer, $34 \times 34 \times 3$ for bilayer, $30 \times 30 \times 1$ for trilayer, and $31 \times 31 \times 1$ for quadlayer system. The monolayer of WS$_{2}$, composed of single S-W-S stacking unit, has a threefold rotational axis (c axis). Due to lack of inversion symmetry with respect to the mirror plane (containing the W atoms), the symmetry properties of monolayer and trilayer WS$_{2}$ are described by a space group  187-P6m2 ($D_{3h}^{1}$) compared to bilayer and quadlayer which have a space group 164-P3m1($D_{6h}^{4}$).

\begin{figure*}[t]
\centering
\begin{tabular}{cc}
\includegraphics[width=54mm]{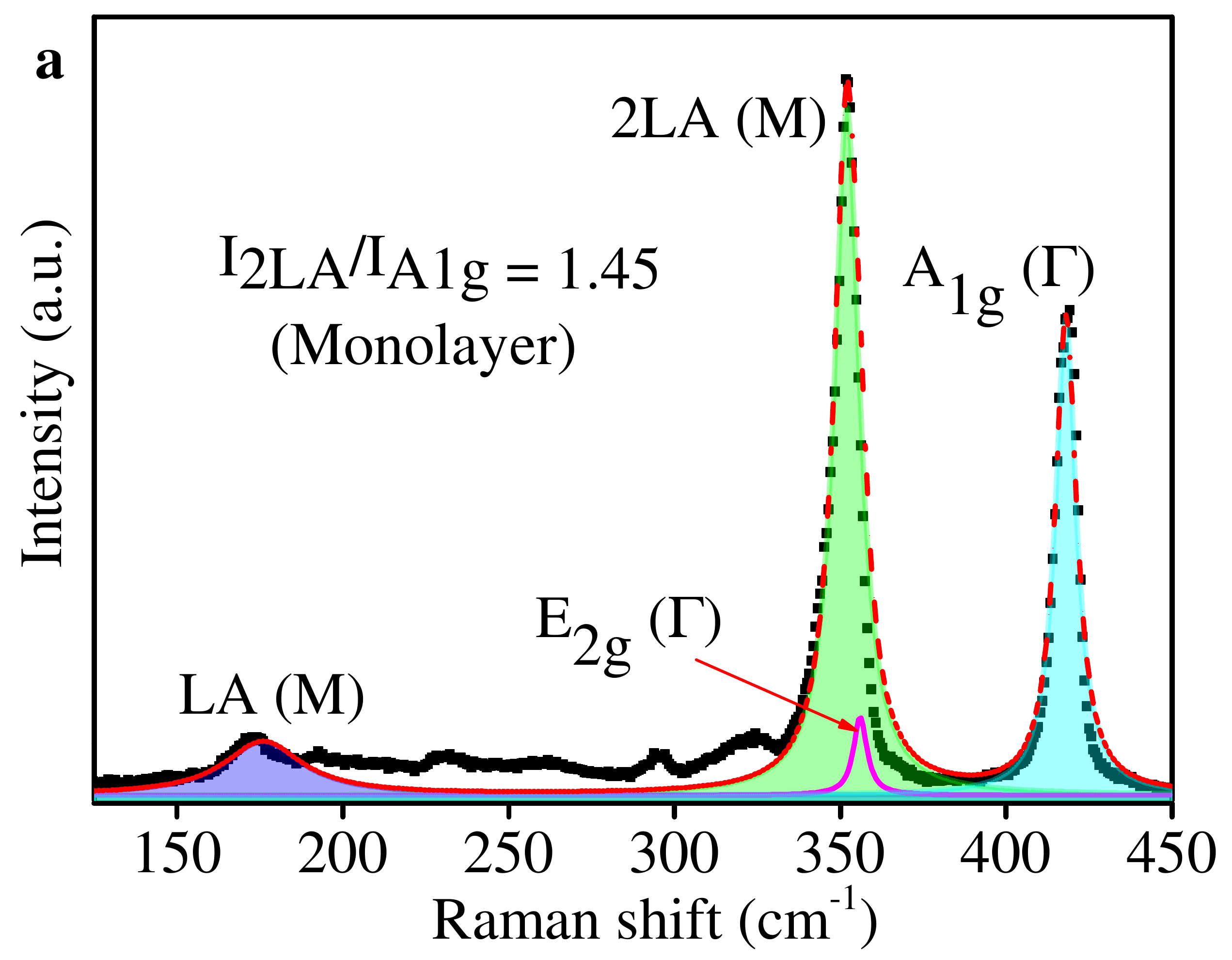}
\includegraphics[width=53mm]{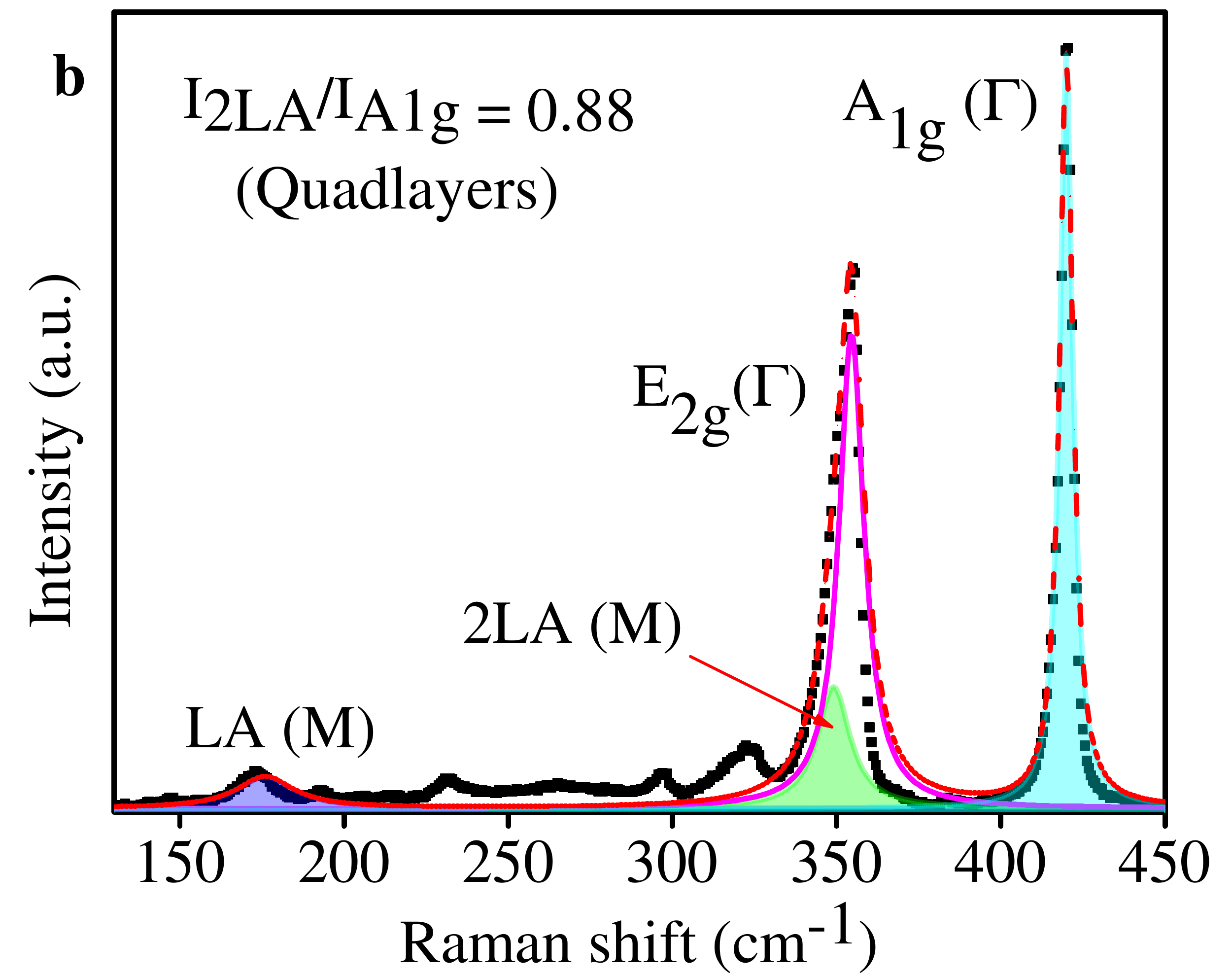}
\includegraphics[width=55mm]{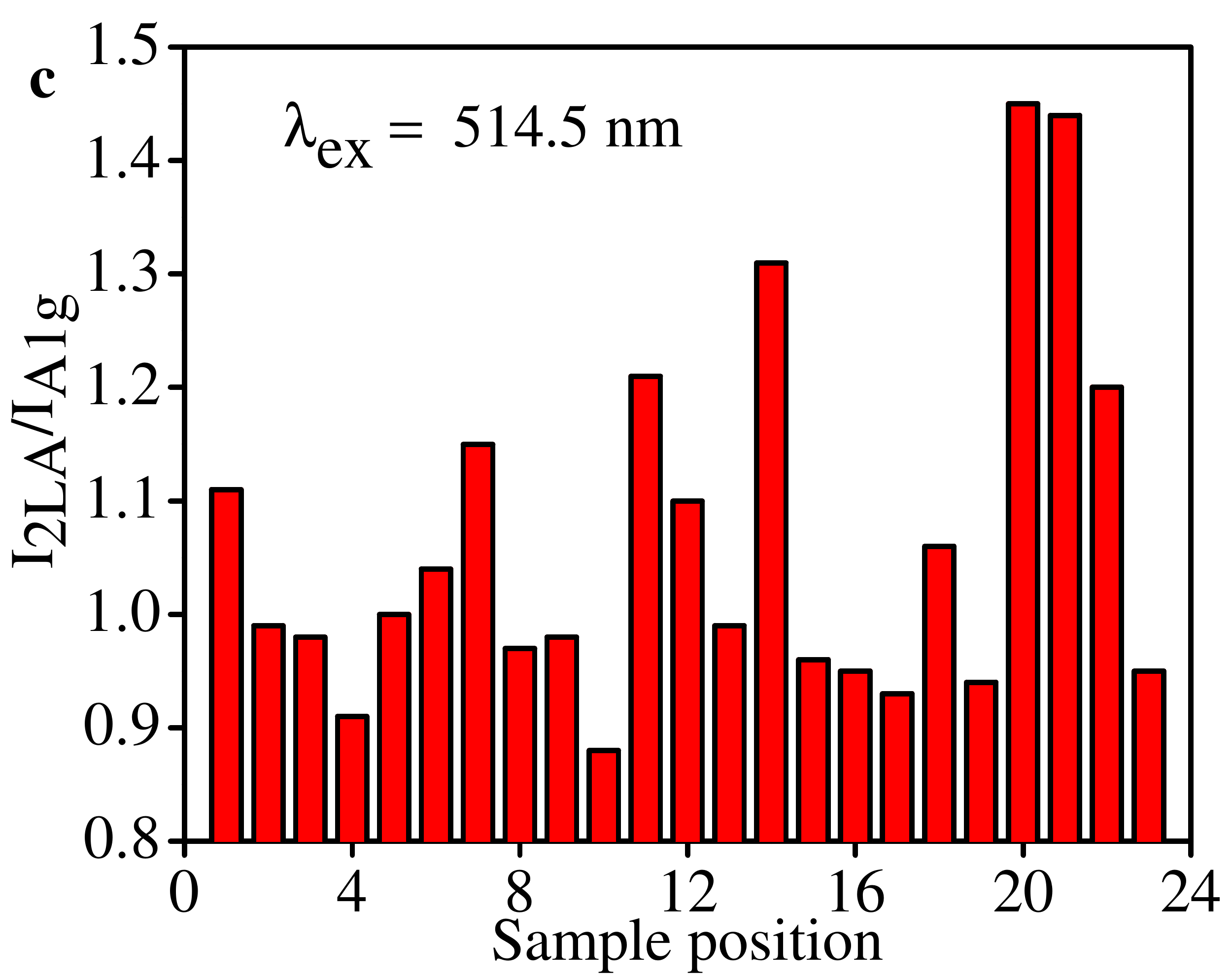}
\end{tabular}
\caption{(Color online) Raman characterizations of exfoliated WS$_{2}$ show the signature of (a) monolayer and (b) quadlayers. The deconvoluated spectra consist of 1st order (blue shaded region) and 2nd order (green shaded region) LA modes along with different vibrational modes (E$_{2g}$ as pink line and A$_{1g}$ as cyan shaded region). (c) The intensity ratios of 2nd order LA and A$_{1g}$ modes at different positions of the sample confirm the number of exfoliated S-W-S layers (mono-to-quad).}
\label{raman}
\end{figure*}

\section{Results}




\subsection{Identification of number of S-W-S Layers in Dispersion} Raman fingerprints can provide the information of different phonon modes that can accurately quantify the number of S-W-S layers in the sample. The Raman intensity ($\lambda_{ex}$ = 514.5 nm) of A$_{1g}$($\Gamma$) phonon mode monotonically diminishes,  whereas the 2nd order longitudinal acoustic (2LA(M)) phonon mode drastically increases  with the decrease of layer numbers as shown in Fig.~\ref{raman}(a) and (b). The intensity ratio (I$_{2LA}$/I$_{A1g}$) of these two Raman modes can unambiguously identify the exact number of S-W-S layers present in the sample~\cite{Raman}. We have spin coated the dispersion of WS$_{2}$ on Si substrates and acquired the Raman signal at several positions. The intensity ratio (I$_{2LA}$/I$_{A1g}$) histogram (Fig.~\ref{raman}(c)) ranging from 0.85 to 1.45 for different sample position confirms that we have successfully synthesized $\sim$ mono-to-quad WS$_{2}$ layers dispersed in DMF solvent.

\subsection{Biexciton Formation} 
\begin{figure*}[htb]
\centering
\begin{tabular}{cc}
\includegraphics[width=42mm]{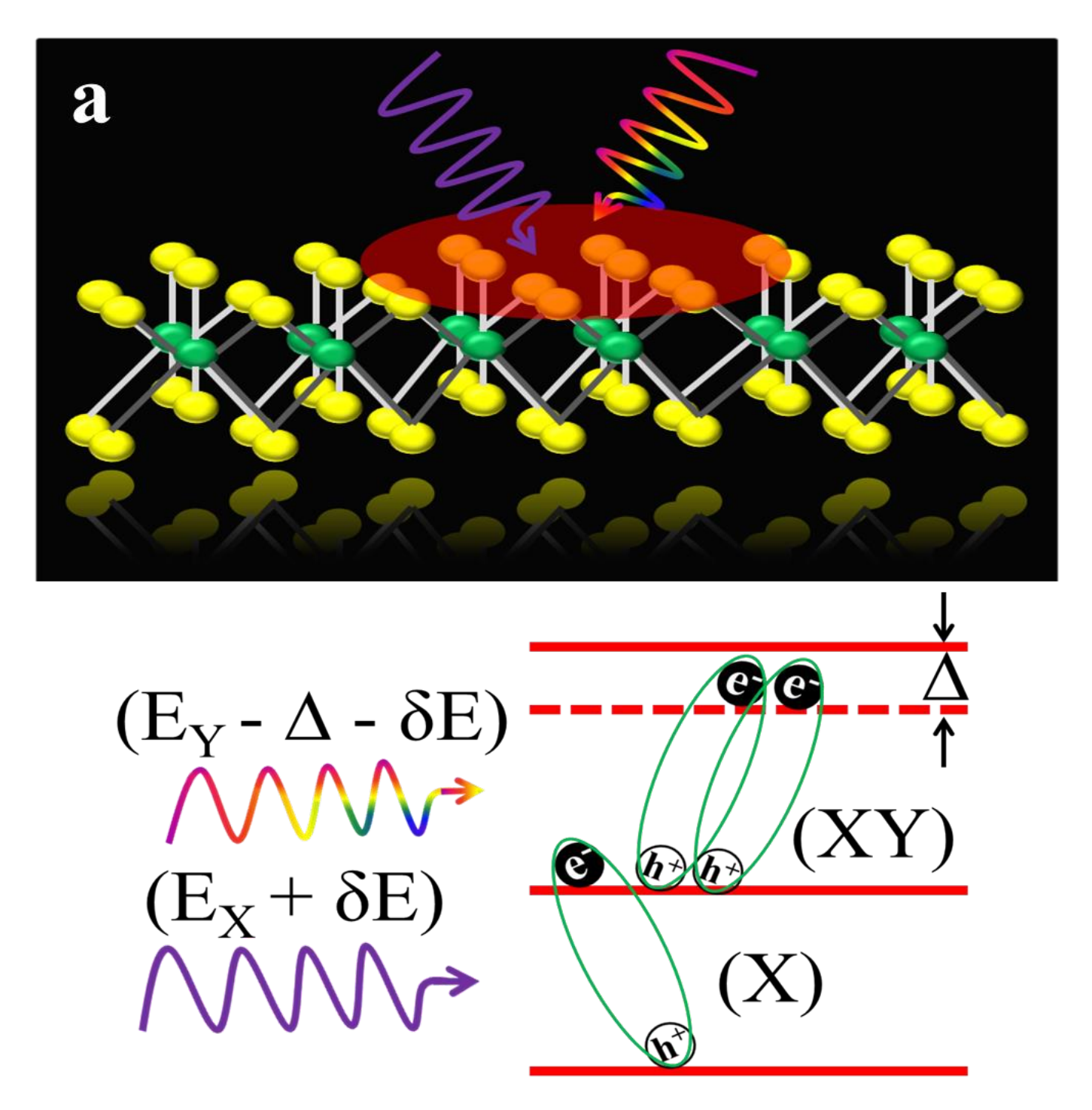}
\includegraphics[width=55mm]{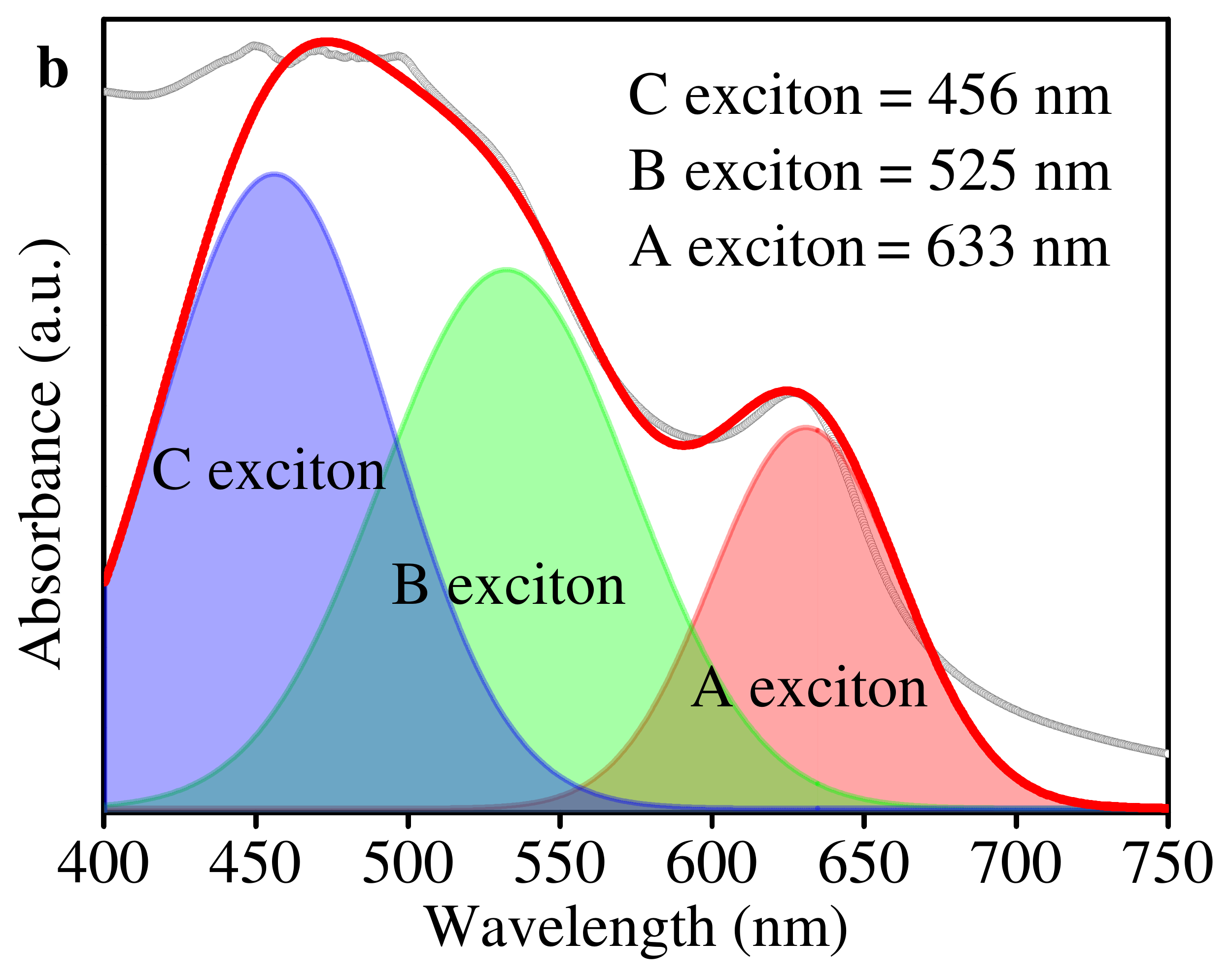}
\includegraphics[width=75mm]{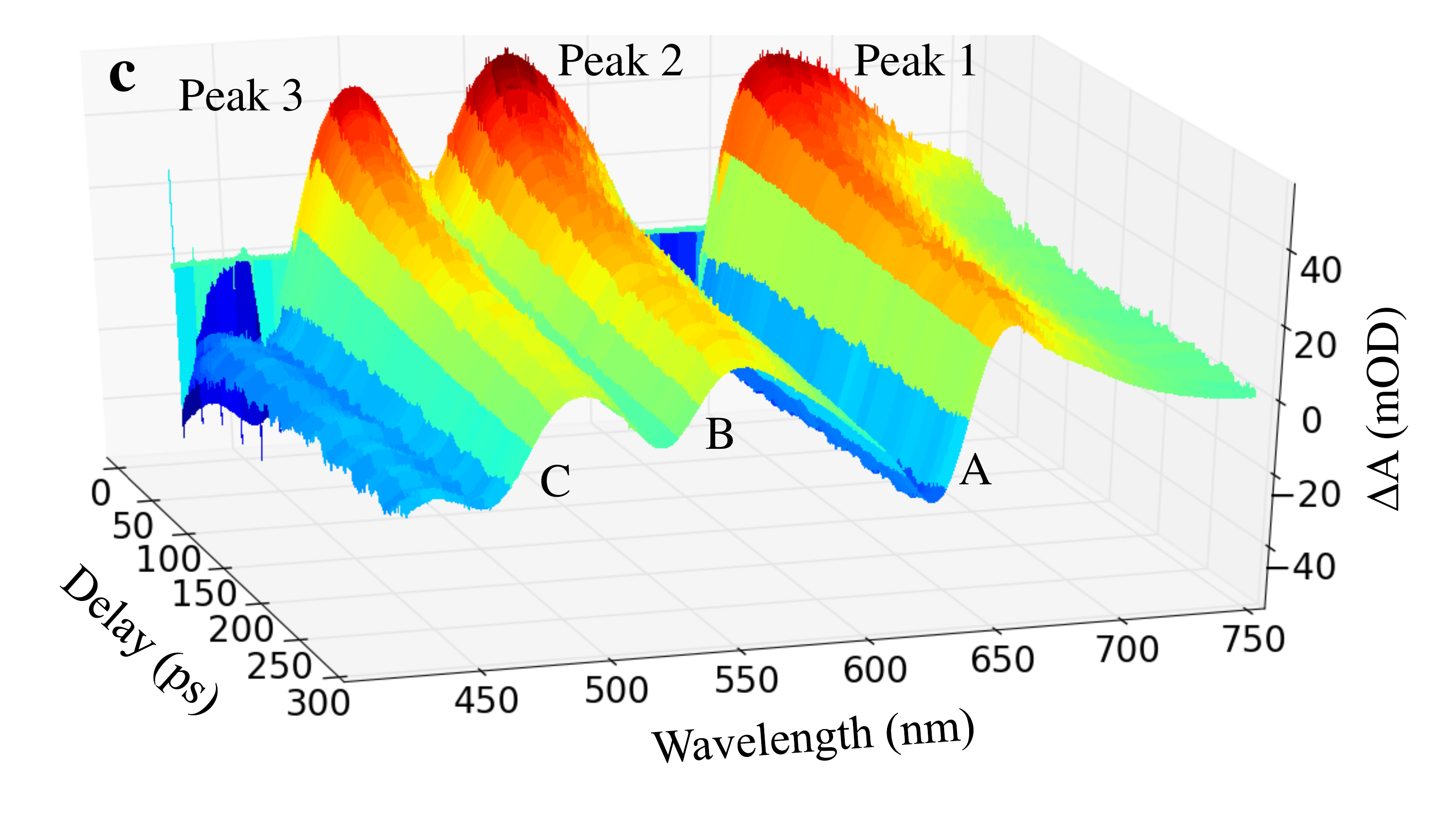}
\end{tabular}
\caption{(Color online) (a) Schematic representation of the formation of biexcitons in layered WS$_{2}$ using two-step pump ($E_{X}$ + $\delta$E)-probe ($E_{Y}$ - $\Delta$ - $\delta$E) excitation process. Here excitons and biexcitons are defined as X and XY respectively. (b) depicts deconvoluted steady state absorption spectrum of WS$_{2}$ dispersion which clearly depicts the formation of A (red shaded region), B (gree shaded region) and C (blue shaded region) excitons. (c) shows the contour map of TA signal for 405 nm, 3 mW pump excitation. Three saturation absorption valleys appear at the positions of steady state excitons (A, B and C), whereas three distinguish pump induced absorption peaks (peak 1, peak 2 and peak 3) appear in the contour map.}
\label{fig1}
\end{figure*}

Unlike conventional semiconductors, biexitons in WS$_{2}$ usually form in a two-steps (ground state-intermediate pseudo excitonic states-biexcitonic state) process governed by cumulative effect of pump and probe photons~\cite{Biexciton B.E. theory, Biexciton B.E. expt ws2} as shown in Fig.~\ref{fig1} (a). In WS$_{2}$, the intermediate pseudo excitonic state appears due to valley splitting~\cite{Paradisanos biexciton WS2, Sie MoS2 pump probe}. As a result of this, different kind of biexcitons form namely AA and BB biexcitons. The biexcitonic features of WS$_{2}$ can be realized directly from TAS spectra in the following manner. The  deconvoluted steady state absorption spectrum is depicted in Fig.~\ref{fig1}(b), whereas the 3D contour plot of the TAS signal as a function of probe wavelength (420-750 nm) and delay (-5 to 300 ps) is depicted in Fig.~\ref{fig1}(c) for mono-to-quad layer WS$_2$ dispersion in DMF solvent. The TAS spectrum shows six major features including three saturated absorption (SA) peaks and three excited state absorption (ESA) peaks along with a pump (405 nm, 3 mW) induced ground state bleaching (GSB). The saturation absorption peaks at three different positions, A (456 nm), B (525 nm), and C (633 nm) corroborate with the deconvoluted steady state absorption peaks of the sample. To distinguish the biexcitonic features from ESA, TAS spectra are deconvoluted using multiple gaussian fit as shown in Fig.~\ref{fig3}(a)-(d). Six gaussian peaks are used to reproduce the complete TAS spectra at different probe delays where the center position of negative gaussian peaks are kept fixed at different wavelengths of steady state excitonic absorption peaks (A, B and C). The positive $\Delta$A features as depicted in Fig.~\ref{fig1}(c) at two different positions 659 nm (peak 1) and 563 nm (peak 2), calculated from deconvoluted peak of TAS signal (Fig.~\ref{fig3}(a)-(d)) suggest the biexciton formation. The broadening effect of these peaks are attributed to intervalley biexcitonic absorptions~\cite{Sie MoS2 pump probe}. In our experiment, the off-resonant deep conduction band excitation (405 nm or 3 eV) by the pump pulse (h$\nu$ $>$ E$_{A}$) imparts an excess energy per exciton ($\delta E$), where E$_{A}$ is the energy of the exciton, leads to the immediate formation of hot exciton gas. The biexciton energy can now be expressed as,
\begin{eqnarray}
E_{XY} = (E_{X} + \delta E)_{pump} + (E_{Y}-\Delta -\delta E)_{probe}
\label{e1}
\end{eqnarray}
where $E_{X}$ and $E_{Y}$ are the energy of the X and Y excitons respectively and $\Delta$ is the biexciton binding energy~\cite{Sie MoS2 pump probe}. Thus the biexcitonic peaks are shifted from the corresponding excitonic peak with a $\Delta$ amount of energy. On the other hand, the origin of peak 3 (503 nm) as shown in Fig.~\ref{fig1}(c) is possibly different, which seems to be just another excited state absorption (ESA) as there is no corresponding steady state absorption (Fig.~\ref{fig1}(b)).


\subsection{Exciton cooling process} 
\begin{figure}[t]
\centering
\begin{tabular}{cc}
\includegraphics[width=43mm]{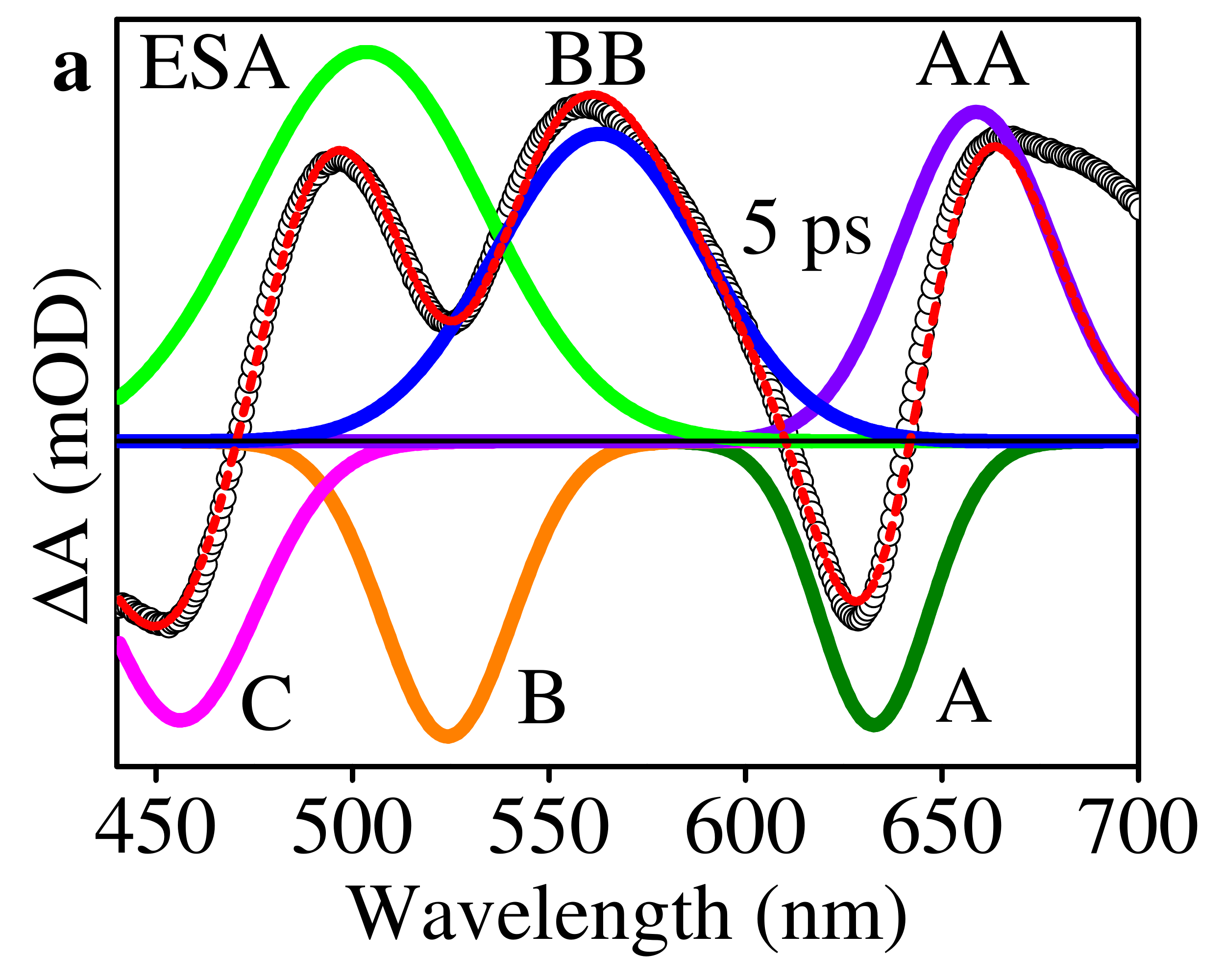}
\includegraphics[width=43mm]{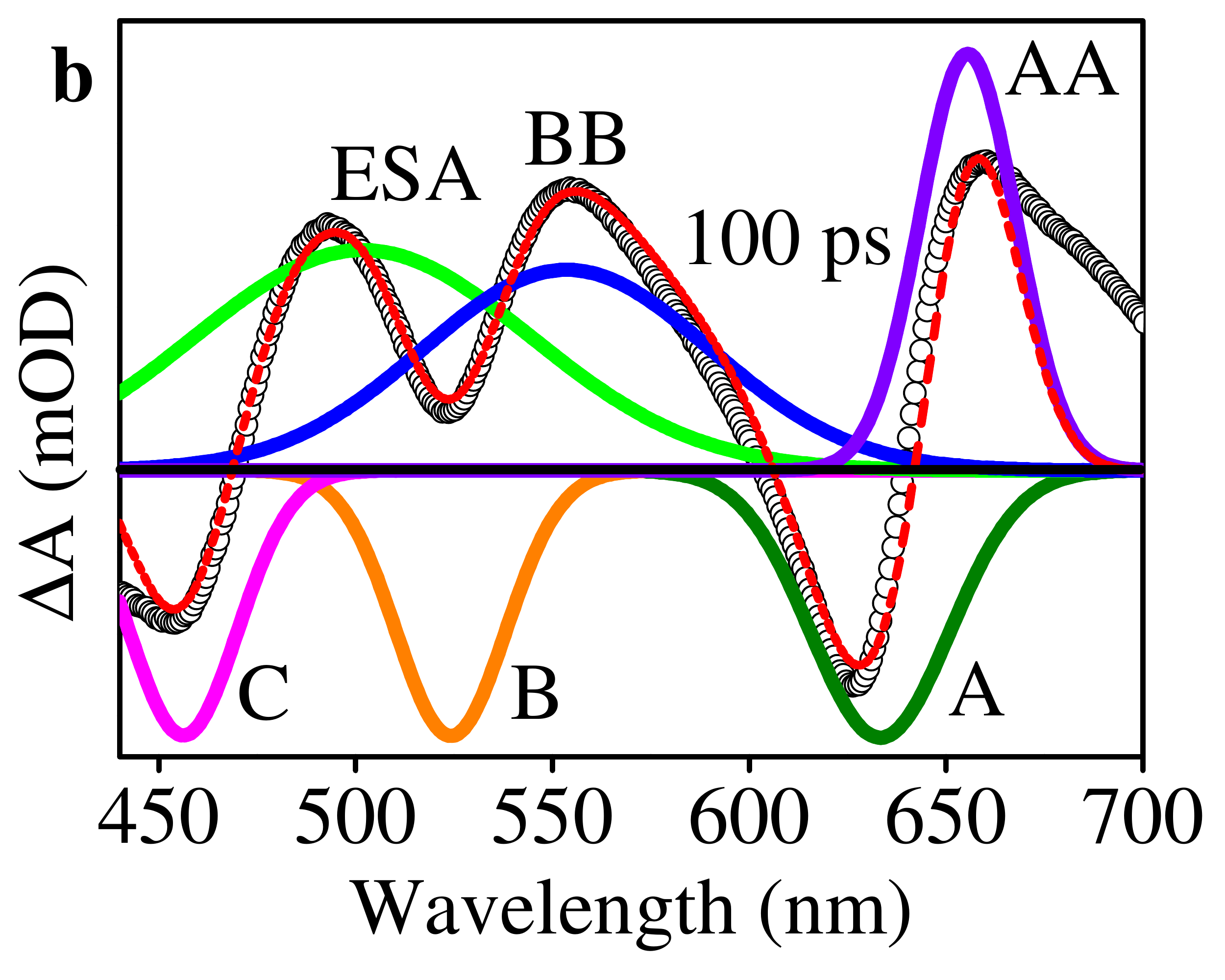}\\
\includegraphics[width=43mm]{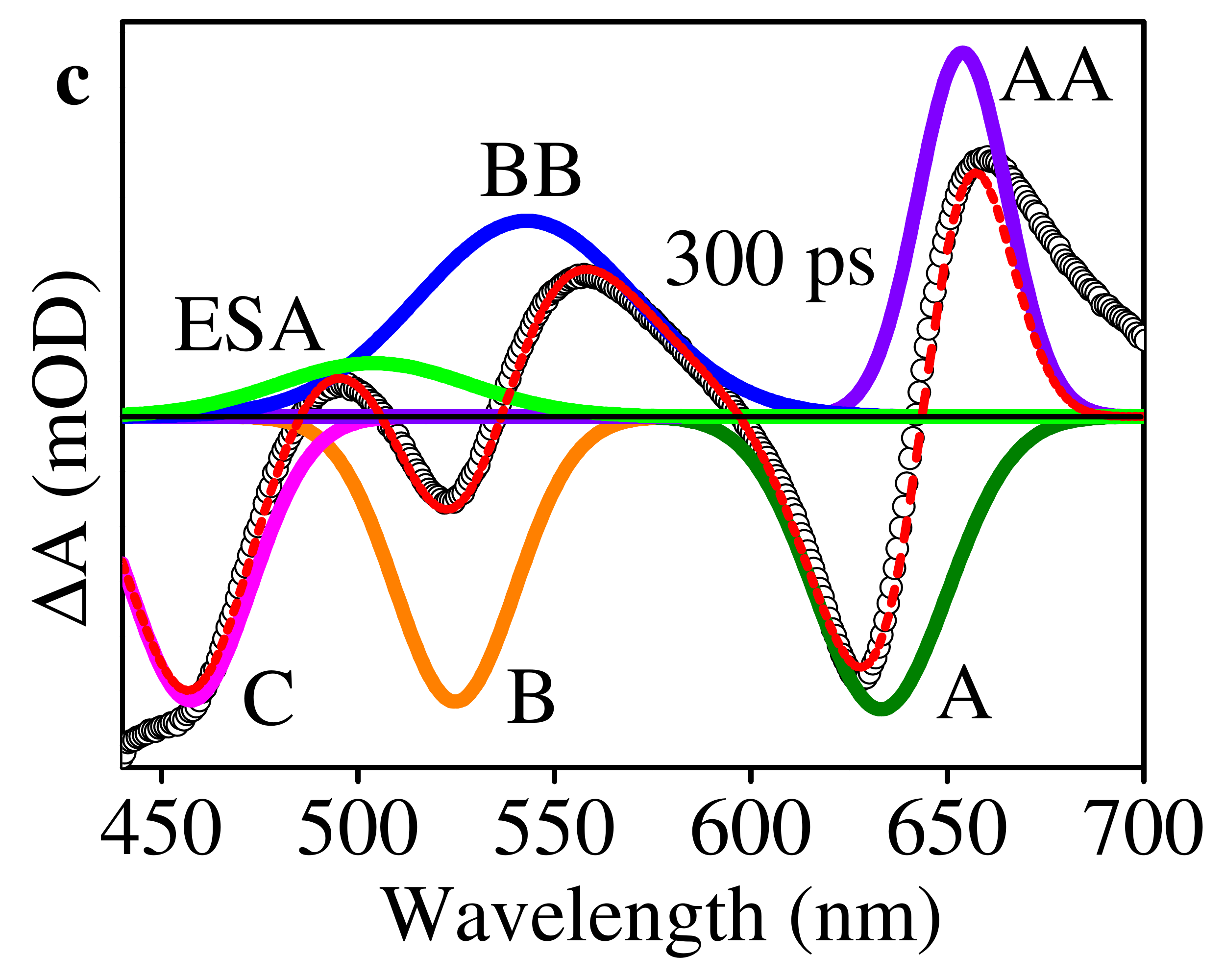}
\includegraphics[width=43mm]{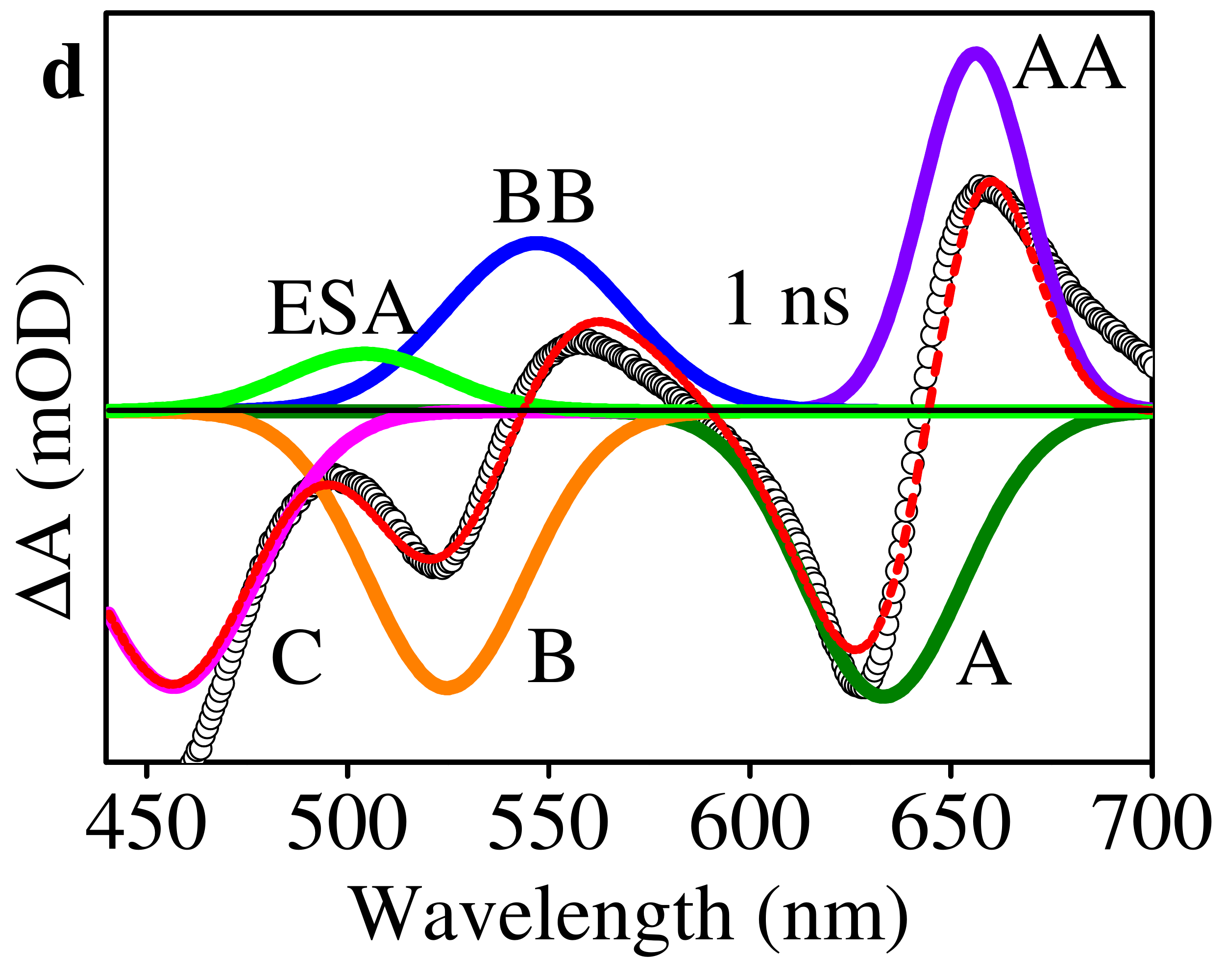}
\end{tabular}
\caption{(Color online) (a)-(d) show time resolved TA spectra of mono-to-quad layer WS$_{2}$ dispersion. The time delay has been varied from 5 ps to 1 ns (1000 ps). All four spectra show six major features containing three saturation absorption (C (pink line), B (orange line) and A (olive line)), two biexcitonic absorption (BB (blue line) and AA (violet line)) and one excited state absorption (green line).}
\label{fig3}
\end{figure}

We have acquired the transient absorption spectra for different probe time delays for further elaboration of the biexciton formation. The TAS signals at different time delays (5 ps, 100 ps, 300 ps and 1 ns) along with deconvoluted peaks are depicted in Fig.~\ref{fig3}(a)-(d). It is clearly seen from the Fig.~\ref{fig3}(a)-(d) that the position of negative peaks (A, B and C) originating from saturation absorption (SA) remain fixed with delay. In presence of pump, the photo-bleached ground states are unable to absorb the probe pulse. As a result, probe directly transmits to detector and position of the negative peaks remain unchanged. On the other hand, we have found that ESA continuously decreases from 5 ps to 300 ps and gets saturated later on, as shown in Fig.~\ref{fig3}(a)-(d). Relative enhancement of oscillator strength of different blue-shifted biexcitonic peaks with respect to ESA are observed with increase in probe delay. Here, the shift is driven by exciton cooling process~\cite{Schmitt-Rink cold exciton}. The measured probe delay dependent blue-shift for biexcitonic peak (containing BB) is $\sim$ 90 meV~\cite{Hulin blueshift} from 5 ps to 300 ps which remains unchanged up to 1 ns. The hot exciton gas gradually relaxes into the cold exciton gas via releasing energy ($\delta E_{probe}$) to the WS$_2$ lattice and DMF solvent. As a result, during the formation of cold exciton gas, the energy of the biexciton population ($E_{XY}$) shifts into lower wavelengths with increasing delay. Using these cooling process, we have also calculated the biexciton binding energies (Tab.~\ref{Tab1}) as $\sim$ 69 meV (AA) and 66 meV (BB) from the difference between the blue shifted biexcitonic peaks and corresponding excitonic (SA) peaks for 1 ns delay using Eq.~\ref{e1} which closely matches with previous theoretical and experimental reports on WS$_2$~\cite{Biexciton B.E. theory, Biexciton B.E. expt ws2, Paradisanos biexciton WS2}. 

\begin{table}
\begin{center} 
\begin{tabular}{| p{1.5 cm} | p{1.5 cm}|} 
\hline 
Biexciton & $\Delta$ (meV) \\ 
\hline\hline 
AA & 69 \\ 
\hline
BB & 66 \\
\hline
\end{tabular}
\end{center} 
\caption{Binding energies of AA and BB biexcitons in layered WS$_{2}$} 
\label{Tab1}
\end{table}

\begin{figure}[t]
\centering
\begin{tabular}{cc}
\includegraphics[width=43mm]{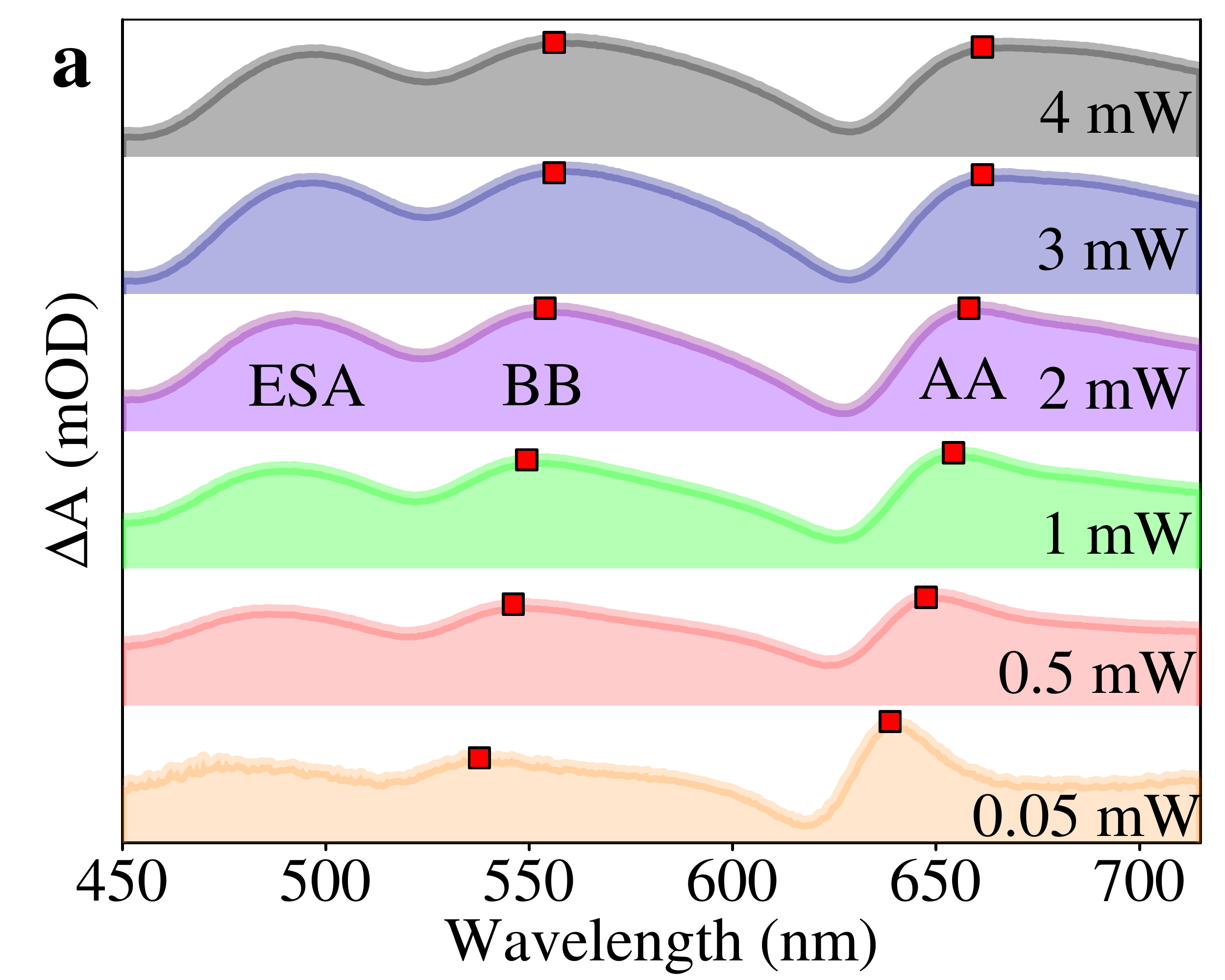}
\includegraphics[width=43mm]{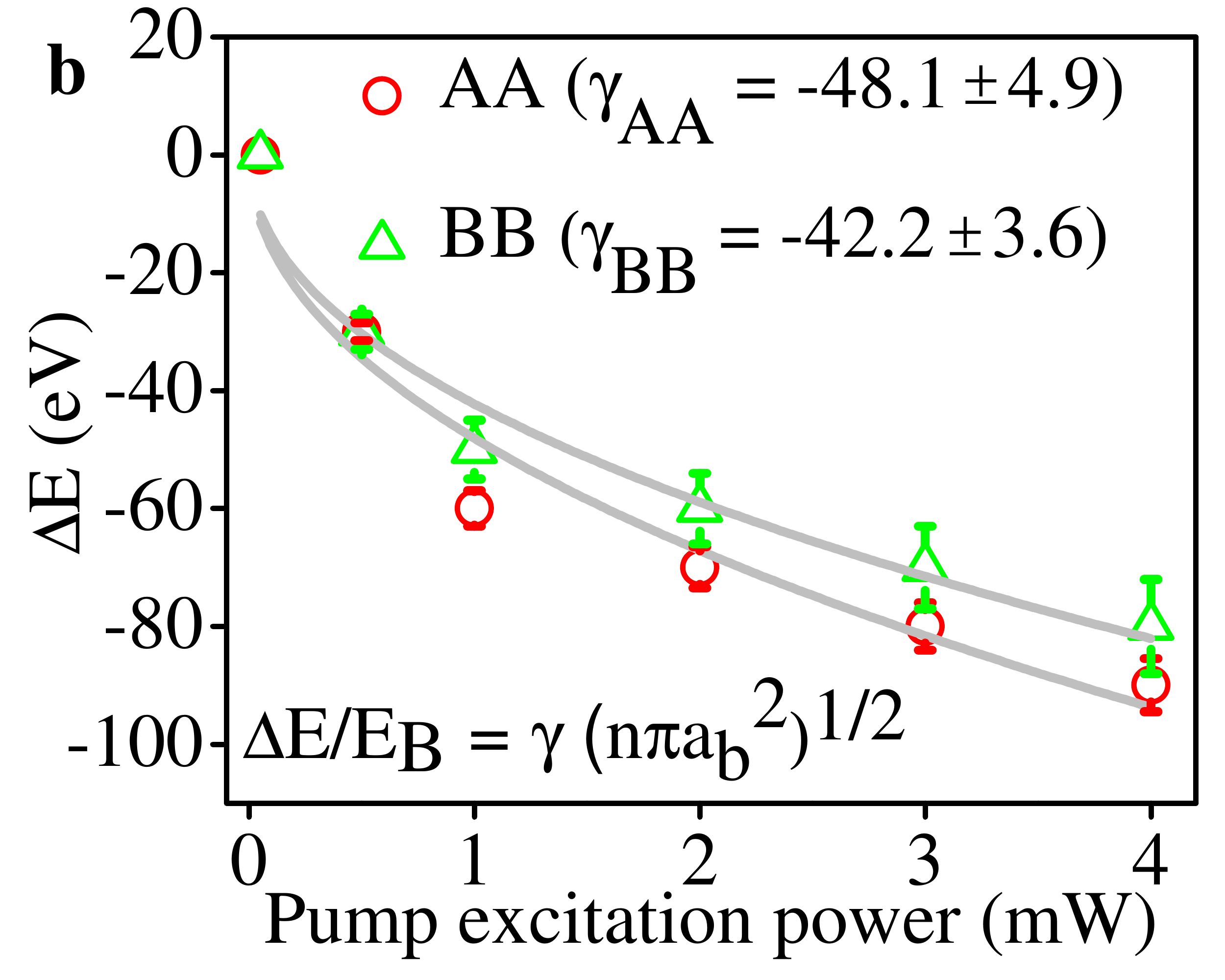}
\end{tabular}
\caption{(Color online) (a) shows pump (405 nm, 3 eV) power dependent (power increases from bottom to top) TA signals for 5 ps probe delay. The biexcitonic peaks are red shifted (red boxes) with increasing power up to 3 mW, whereas the peak positions of ESA remain unchanged. (b) shows variation of biexcitonic red shift ($\Delta$E) with pump power for AA (red circle) and BB (green triangle) biexcitons. The data has been fitted by Eq.~\ref{e2}.}
\label{fig4}
\end{figure}

\begin{figure*}[t]
\centering
\begin{tabular}{cc}
\includegraphics[width=45mm]{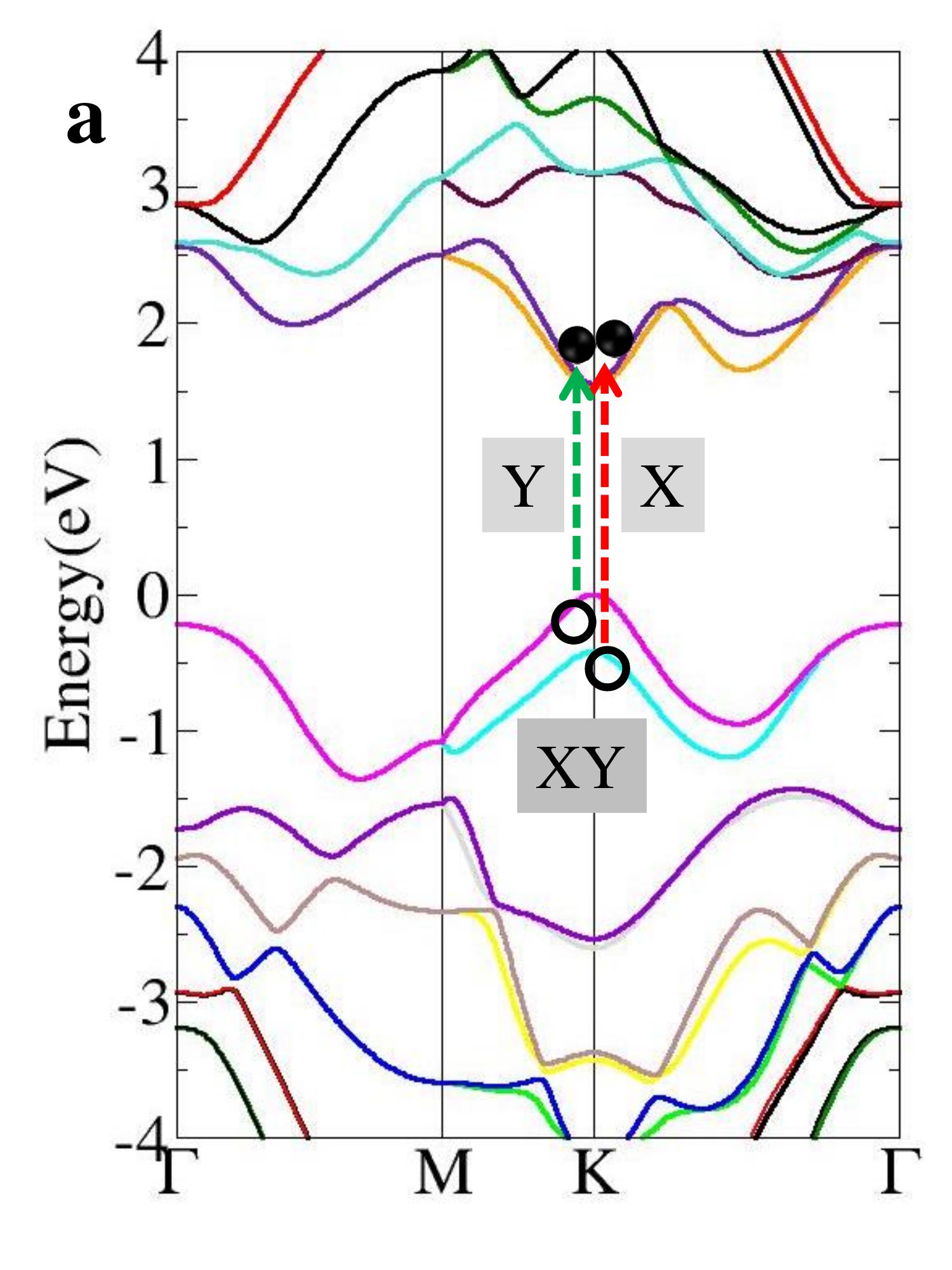}
\includegraphics[width=45mm]{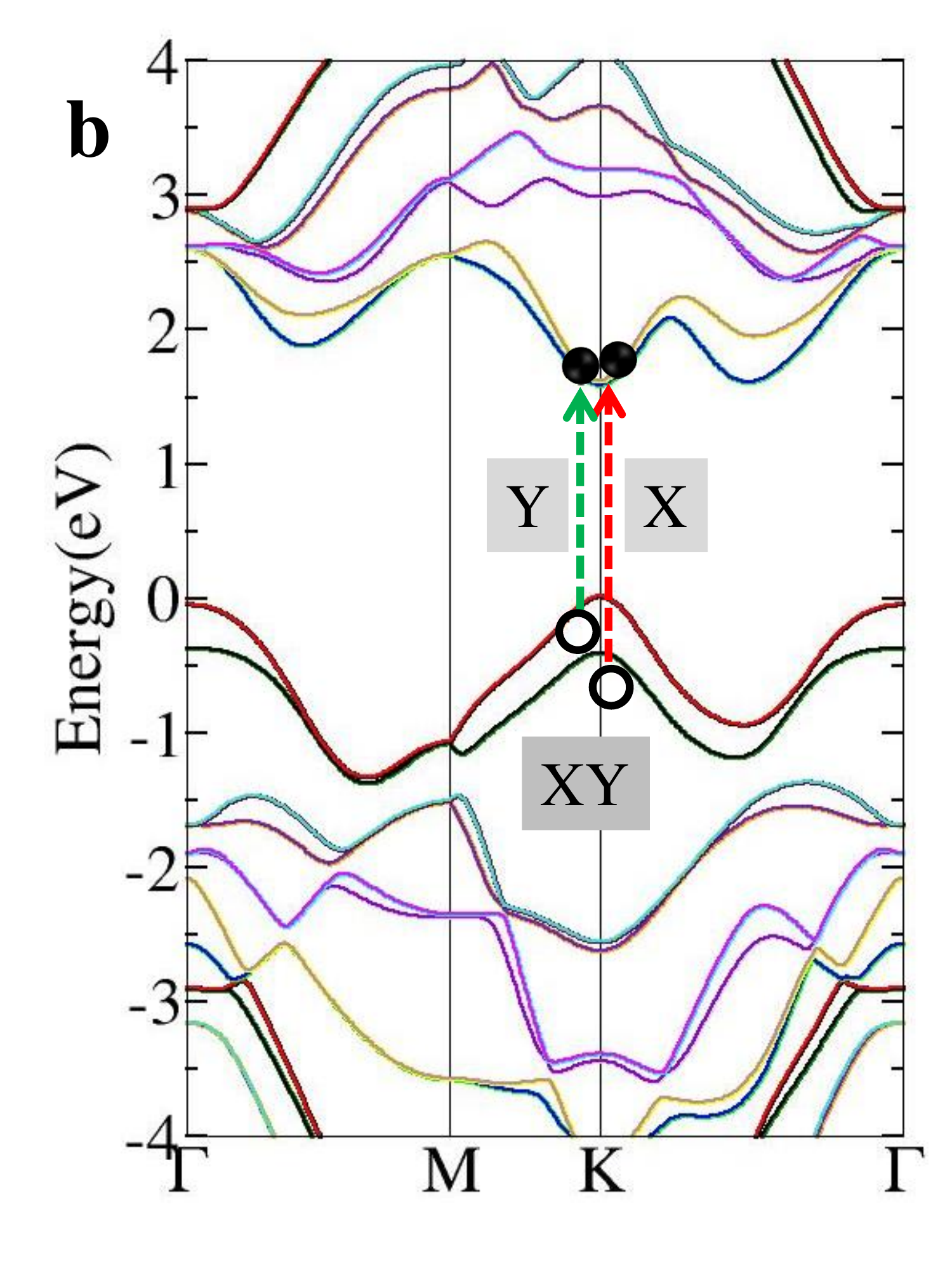}
\includegraphics[width=45mm]{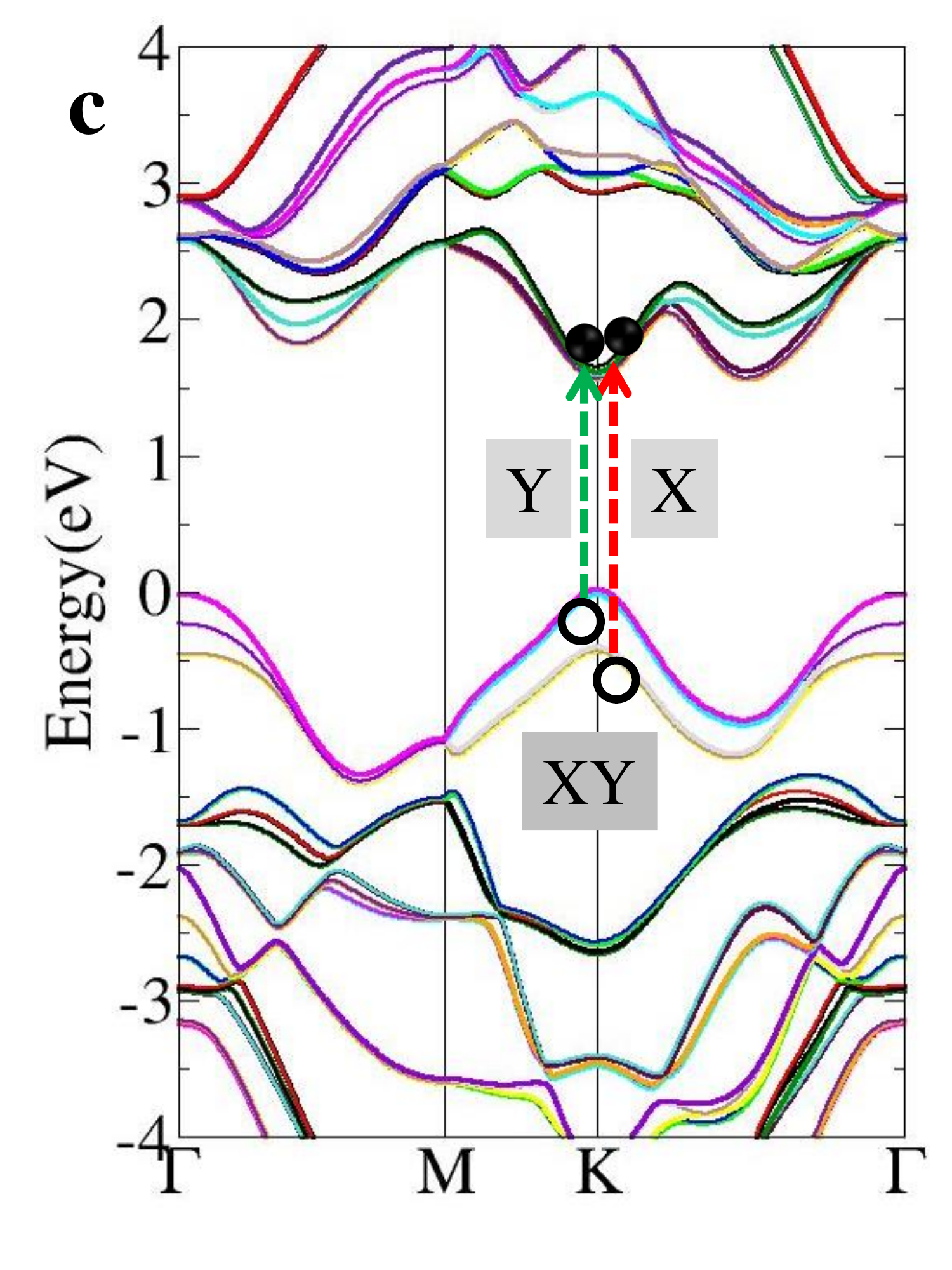}
\includegraphics[width=45mm]{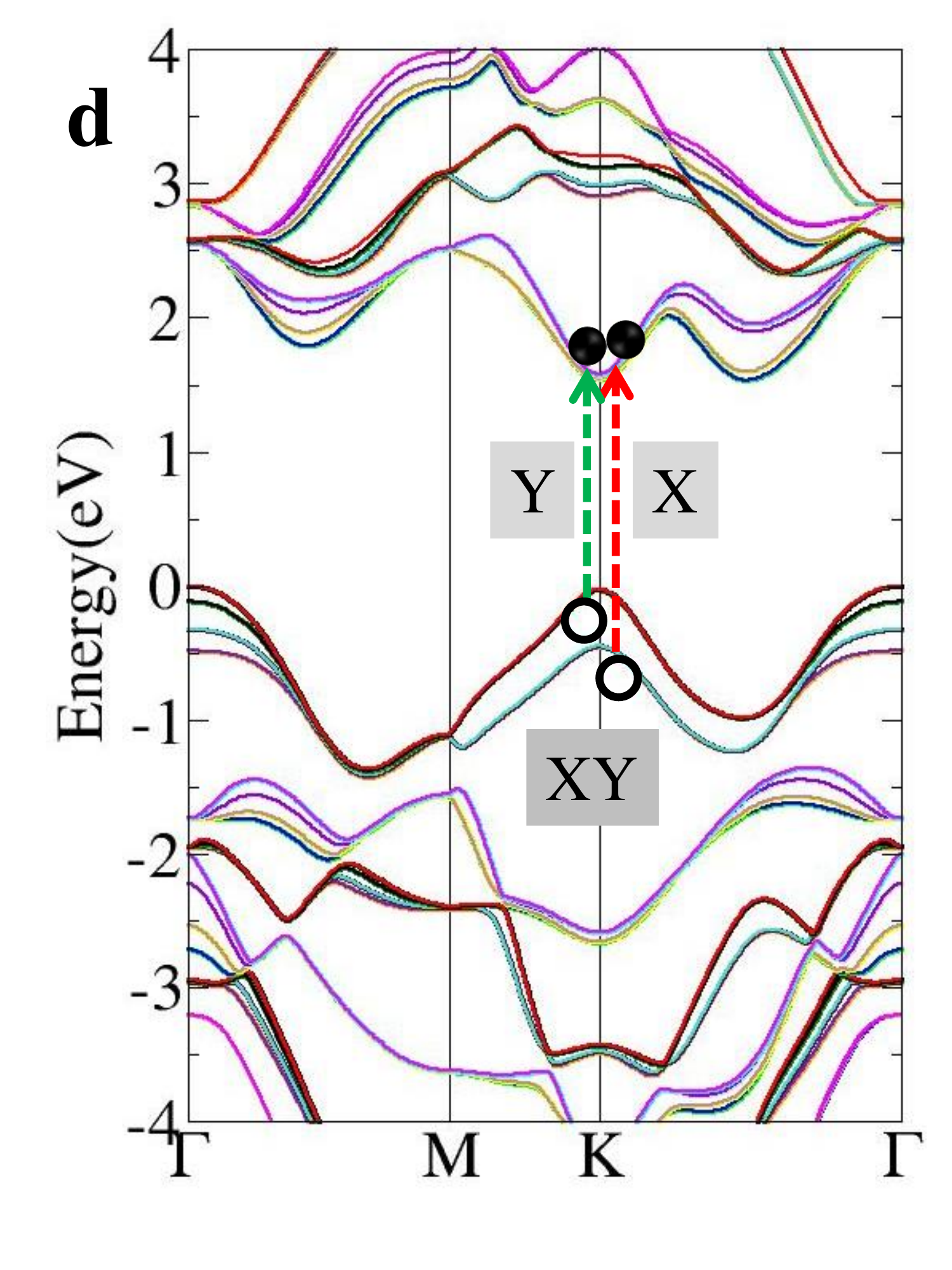}
\end{tabular}
\caption{(Color online) Electronic band structures calculated for the WS$_{2}$ (a) monolayer, (b) bilayer, (c) trilayer and (d) quadlayer systems with inclusion of the spin-orbit interaction. It is clear that the valence band splittings at K point are nearly independent of the number of layers. The biexciton formation via two-step pump-probe process is schematically shown for all four cases.}
\label{fig2}
\end{figure*}


\subsection{Many-body interactions} We find that biexcitons are formed at room temperature for the studied system due to strong many body interactions which results into red-shift of biexcitonic peaks with increasing pump power. To verify this strange behavior, we acquire series of $\Delta$A (Fig.~\ref{fig4}(a)) for different pump (405 nm, 3 eV) photon energies at 5 ps probe delay. The 5 ps time delay has been chosen, as the intervalley spin relaxation occurs in a time scale within a few hundreds of femtoseconds~\cite{Mai spin flipflop}. As we are interested in biexciton formation process, we will focus on the 500-700 nm probe range of TAS signal. When the pump power is increased gradually from 50 $\mu$W to 4 mW (0.03 to 2.4 GW/cm$^{2}$), the positive peaks shift to lower energies and develop a low-energy shoulder as shown in Fig.~\ref{fig4}. With increasing pump power, the number of biexciton formation enhances due to increase of exciton population, results into red shift of the biexcitonic peaks~\cite{Sie MoS2 pump probe, Sie WS2 shift}. The calculated red-shift of these two peaks are $\sim$ 80 meV (BB) and $\sim$ 90 meV (AA) for 3 mW pump power, with no additional shift up to 4 mW. At higher pump powers, the many body interactions get additionally screened by the photoexcited carriers and modify exchange-correlation energies. On the other hand, the electronic band structure becomes renormalized which causes the reduction of exciton binding energy as well as enhancement in absorption of pump photons~\cite{Ferrari Band renorm., Sie MoS2 pump probe, Aleithan MoS2 band map pumpprobe}. Since, the Coulomb interactions will tend to recombine the e-h pair and reduction of exciton binding energy due to band renormalization as well as the modified dielectric screening due to the distribution of S-W-S layer numbers will try to make them free~\cite{Dielectric theory1, Dielectric theory2}. The trade off between these two effects often results in a net energy shift ($\Delta$E) of the biexcitonic peak. The red shift ($\Delta E$) of the biexcitonic peak can be calculated by the following equation,
\begin{eqnarray}
\frac{\Delta E}{E_{B}}=\gamma (n\pi a_{b}^{2})^{k}
\label{e2}
\end{eqnarray}
where, E$_{B}$ is the exciton binding energy, $a_{b}$ is the exciton Bohr radius, k and $\gamma$ are dimensionless factor. In Eq.~(\ref{e2}), $\Delta$E obeys a power-law nature with pump power (n), until the interexciton distance ($\propto\frac{1}{\sqrt{n}}$) approaches the exciton Bohr radius, where the Weiner-Mott excitonic transition is expected~\cite{Wang exciton, Schmitt-Rink cold exciton}. It has been found that the short-range interactions will dominate for $\Delta E \propto \sqrt{n}$ whereas the long-range interaction becomes stronger for $\Delta E \propto n$~\cite{Sie MoS2 pump probe}. We find that the experimental results are best fitted with $k = 1/2$ and corresponding $\gamma$ values are $-48.1 \pm 4.9$ and $-42.2\pm3.6$ for AA and BB, respectively as shown in Fig.~\ref{fig4}(b). So, we are in such a situation where short-range interactions are dominant and the excitons are found to be Weiner-Mott excitons~\cite{Sie MoS2 pump probe, Wang exciton}.

\subsection{Origin of saturation absorption} The dispersion, used in this experiment, has the information of mono-to-quad layer of WS$_{2}$ as confirmed by Raman analysis. Therefore, it would be worth knowing whether SA peaks in TAS spectrum appears due to one particular layer or a combination of all layers of the WS$_2$ dispersion. To investigate the origin of SA peaks extracted from the pump-probe measurement, we have computed electronic band structure of monolayer, bilayer, trilayer and quadlayer WS$_{2}$ computed using first-principles calculations are shown in Fig.~\ref{fig2}(a)-(d). Our calculations reveal that the monolayer WS$_{2}$ is a direct-band-gap semiconductor with the valence band maxima (VBM) and conduction band minima (CBM) at K and K$^{\prime}$ point of the Brillouin zone, in contrast to the indirect band gap of the bilayer, trilayer and quadlayer WS$_{2}$, also observed in angle-resolved photoelectron spectroscopy experiment~\cite{Klein, Swasti, Zhu, Lu spin valley}. In all four cases, the energy bands split at all k points except some points with special symmetry in the Brillouin zone due to the atomic spin-orbit coupling. The spin-orbit interaction breaks the spin degeneracy of the valence and conduction bands along the $\Gamma$-K-M direction. It is found that the splitting in VB is largest at the K point of the Brillouin zone whereas the CBM is also at K, as depicted in Fig.~\ref{fig2}. The conduction band and valence band of WS$_{2}$ are composed of mainly d orbitals of W and relatively small admixture of p orbitals of S, which constitute spin splitting correspond to the K and $K^{\prime}$ valleys~\cite{Mai spin flipflop, Dou MoS2 intervalley}. The lifting of degeneracy in VB appears due to spin-orbit coupling in the E-K diagram for all four cases as shown in Fig.~\ref{fig2}(a)-(d).   

From our calculations, it is clear that the spin orbit interaction invoked valence band splitting, leads to energetically well-separated excitonic transitions which give rise to two distinct low-energy peaks at position A and B in the TAS spectrum as shown in Fig.~\ref{fig1}(c). In this case, the pump pulse populates different valleys of conduction band. The electron population of one particular valley spreads over other valley due to inter-valley spin relaxation within a few hundred of femtoseconds, leads to the formation of dark excitons~\cite{Mai spin flipflop}. Now, the SA peaks which appear mainly due to bright excitonic states, may also be present due to Pauli blocking of the dark excitonic states in the unpumped valleys in absence of pump ~\cite{Ye dark exciton WS2, Zhou dark exciton}. 

On the other hand, the origin of C peak is different from A and B peaks. It is strongly believed that this peak appears due to steady state absorption from deep VB to CB~\cite{Chowdhury exfoliation}. The detailed theoretical investigations reveal that the C peak actually originates from the locally parallel band along $\Gamma$-K direction which is also reffered as band nesting (BN)~\cite{Carvalho BN}. The probability of excitonic transitions are governed by the joint density of states (JDOS)~\cite{Liang JDOS} which can be written as
\begin{eqnarray}
\rho^{J} (\omega)=\frac{1}{(2\pi)^{2}} \int_{S(\omega)} \frac{dS}{|\nabla_{k}(E_{c}-E_{v})|}
\end{eqnarray}
where $E_{c}$ and $E_{v}$ are energies of CB and VB respectively. It is clearly seen from the figure that $E_{c}$ and $E_{v}$ at high symmetrical K and $K^{\prime}$ points follow $\nabla_{k}E_{c}=\nabla_{k}E_{v}=0$ condition which leads to Van Hove singularity (VHS) in JDOS. As the transition probability at these K and $K^{\prime}$ points becomes maximum due to VHS, the saturation absorption appears at A and B peak positions. On the other hand, the condition $\nabla_{k}(E_{c}-E_{v}) = 0$ along $\Gamma - K$ direction with $|\nabla_{k}E_{c}| \sim |\nabla_{k}E_{v}| > 0$ indicates band nesting, gives rise to SA at C peak in TAS signal due to the singularities in JDOS.   

\begin{table*}
\begin{center} 
\begin{tabular}{| p{2.5cm} || p{2.5cm} | p{2.0cm} | p{2.2cm} | p{2.0cm} | p{2.2cm}| p{2.0cm} |} 
\hline 
Probe wavelength & C exciton & ESA & B exciton & BB biexciton & A exciton & AA biexciton \\ 
\hline\hline 
T1 (ps) & 1.36 $\pm$ 0.06 & 2.37 $\pm$ 0.16 & 1.34 $\pm$ 0.05 & 1.64 $\pm$ 0.08 & 1.19 $\pm$ 0.03 & 1.52 $\pm$ 0.04 \\ 
\hline
T2 (ps) & 21.3 $\pm$ 1.9 & 35.9 $\pm$ 4.1 & 34.2 $\pm$ 20.2 & 42.3 $\pm$ 4.1 & 94.9 $\pm$ 73.4 & 54.4 $\pm$ 8.1 \\
\hline
T3 (ps) & 1713.3 $\pm$ 263.0 & 297.4 $\pm$ 21.9 & 95.4 $\pm$ 16.4 & 503.9 $\pm$ 38.6 & 1494.9 $\pm$ 851.0 & 773.8 $\pm$ 60.8 \\ 
\hline 
\end{tabular}
\end{center} 
\caption{Non-radiative and radiative lifetimes of different excitons and biexcitons.} 
\label{tab2}
\end{table*}

\begin{figure}[htb]
\begin{center}
\epsfig{file=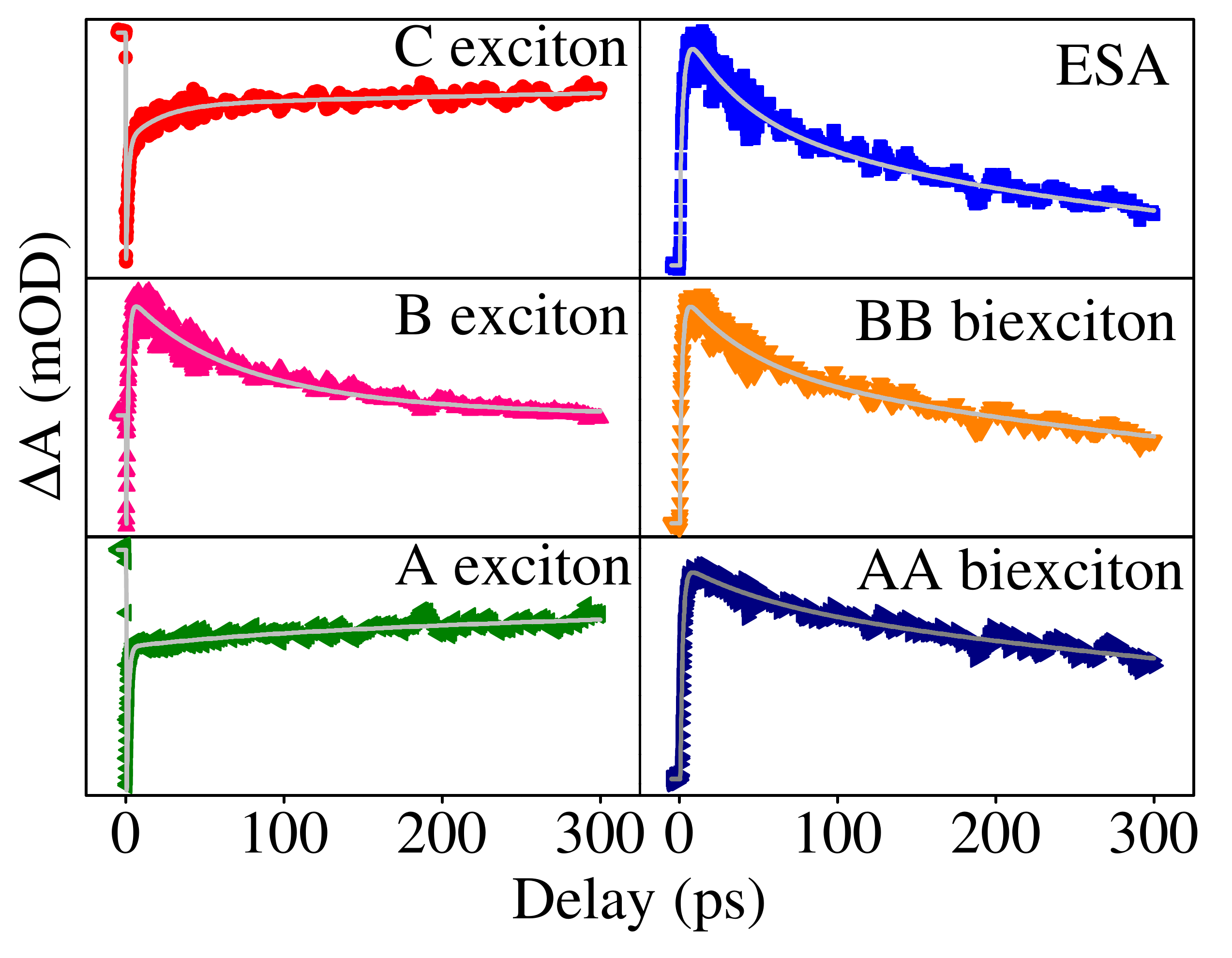,trim=0.0in 0.05in 0.0in 0.05in,clip=true, width=85mm}\vspace{0em}
\caption{(Color online) The ultrafast non-radiative and radiative lifetimes of the excitons and biexcitons are calculated using 3$^{rd}$ exponentially fitted TAS spectra with 200 fs instrument response function (IRF). In this figure, we have represented A, B and C excitons with olive, pink and red markers respectively. ESA is represented as blue marker whereas AA and BB biexcitons are denoted with navy blue and orange markers respectively. Grey lines are used as 3$^{rd}$ exponential fitting.}
\label{fig5}
\end{center}
\end{figure}

\subsection{Charge carrier decay dynamics} 
To investigate the excitonic quasiparticles decay dynamics, we have fitted the TAS signal with a multi-exponential decay equation containing three time constants (T$_{1}$-T$_{3}$) along with an instrument response funtion (IRF) fixed at 200 fs for different probe wavelengths as shown in Fig.~\ref{fig5}. Typically, for TMDs, the T$_{1}$ component is consigned as Auger scattering ranging in few picoseconds\cite{Cunningham augerws2}. The $2^{nd}$ and $3^{rd}$ decay components ($T_{2}$ and $T_{3}$) describe the time scale of non-radiative (rapid relaxation due to surface defect states) and radiative (band-to-band transition) recombinations, respectively~\cite{Sun T2, Chowdhury exfoliation2}. The values of all three time constants for all six peaks of the TAS signal are tabulated in Tab.~\ref{tab2} for 405 nm, 3 mW (1.8 GW/cm$^{2}$) pump excitation. Interestingly, we have found that T$_{1}$ and T$_{2}$ i.e. the non-radiative decay components are similar for all kinds of excitons and biexcitons. However, the radiative component i.e. T$_{3}$ is varying depending upon different quasiparticles. We observe that A and C excitons have similar kind of decay characteristics as SAs are resonantly coupled with each other whereas there is an anomaly for T$_{3}$ of B exciton. From the TAS spectra as shown in Fig.~\ref{fig3}, it is evident that the negative saturation absorption valley for B exciton is overwhelmed due to the presence ESA and BB biexciton. Our fitting results in a smaller decay component (T$_{3}$) of $\sim$ 300 ps for ESA, whereas the typical radiative lifetime of exciton (T$_{3}$ of A and C) are coming out to be in the order of nanosecond. Thus, the presence of ESA may explain the reason behind the decreased radiative lifetime of B exciton. On the other hand, in the case of pump induced absorptions, the radiative decay constants of ESA are much smaller than the biexcitonic peaks which are corroborated with the previously discussed time evolution of TAS spectra (Fig.~\ref{fig3}).
\section{Discussion}
In conclusion, we have established the formation and evolution of ultrafast excitonic quasiparticles (excitons and biexcitons) at room temperature and presented their decay dynamics in the solvent (DMF) exfoliated mono-to-quad layer WS$_{2}$ dispersion using optical pump (405 nm) and broadband UV-visible probe (350-750 nm). Transient absorption spectra reveal the existence of excitonic and biexcitonic features in the dispersion similar to a monolayer WS$_{2}$, as predicted by theoretical results. The first principle calculation suggests that the loss of degeneracy in both VB and CB are still prominent in the modulated layer numbers. The blue shift ($\Delta_{BB}$  $\sim$ 90 meV) of biexcitonic peak occurs due to exciton cooling process whereas pump power dependent red shifts ($\Delta_{AA}$  $\sim$ 90 meV and $\Delta_{BB}$  $\sim$ 80 meV) appear due to strong many body interactions among the quasiparticles. The binding energies of AA ($\sim$ 69 meV) and BB ($\sim$ 66 meV) biexcitons extracted experimentally that agree well with the theoretically reported layered WS$_2$~\cite{Biexciton B.E. theory}. We have shown that the excitons act like Weiner-Mott excitons due to the dominance of short-range coulomb interactions and explained the origin of the generation of different excitons (A, B and C) via first-principles calculations as well. Moreover, from time-resolved studies, we have measured the ultrafast non-radiative and radiative lifetimes of different quasiparticles using three exponential decay processes. In summary, the detailed pump-probe spectroscopic investigation of chemically synthesized layered WS$_{2}$ provided a precious  information of the excitonic quasiparticles dynamics, that can be conducive to explain the novel optical properties of this material, which may set a new paradigm for cost-effective large-scale production compatible WS$_{2}$ based device applications.

\section{Acknowledgments}

RKC, SB and SN  acknowledge  MHRD, India for support. RKC acknowledge Mr. Subhojit Jana for his help in calculations. SN thanks Dr. Monodeep Chakraborty for useful discussions. RKC and SB thanks Dr. Simone Peli for his helpful inputs. Authors acknowledge SDGRI-UPM project of IIT Kharagpur for necessary equipment support in Ultrafast Science Lab, Department of Physics, IIT Kharagpur.

\end{document}